\documentclass[12pt]{iopart}
\usepackage{graphicx}

\usepackage{iopams}
\begin{document}

\title[Spin fluctuations probed by  NMR in paramagnetic spinel  LiV$_2$O$_4$]
{Spin fluctuations probed by  NMR in paramagnetic spinel  LiV$_2$O$_4$: a self-consistent renormalization theory}

\author{V  Yushankhai$^{1,2}$,  T  Takimoto $^{1}$  and  P  Thalmeier$^{1}$}

\address{ $^1$Max-Planck-Institut f\"ur Chemische Physik fester Stoffe, D-01187 Dresden, Germany}
\address{ $^2$Joint Institute for Nuclear Research, 141980 Dubna, Russia}

\ead{\mailto{yushan@cpfs.mpg.de}, \mailto{yushankh@theor.jinr.ru}}

\begin{abstract}

Low frequency spin fluctuation dynamics in paramagnetic spinel
LiV$_2$O$_4$, a rare 3$d$-electron heavy fermion system, is
investigated. A parametrized self-consistent renormalization (SCR)
theory of the dominant AFM spin fluctuations is developed and
applied to describe  temperature and pressure dependences of the
low-$T$ nuclear spin-lattice relaxation rate $1/T_1$ in this
material. The experimental data for  $1/T_1$ available down to
$\sim 1$K are well reproduced by the SCR theory, showing the
development of  AFM spin fluctuations as the paramagnetic metal
approaches a magnetic instability under the applied pressure. The
low-$T$ upturn  of   $1/T_1T$ detected below 0.6 K under the
highest applied pressure of 4.74 GPa is explained as the nuclear
spin relaxation effect due to the spin freezing of magnetic
defects unavoidably present in the measured sample of
LiV$_2$O$_4$.
\end{abstract}


\submitto{\JPCM}

\maketitle

\section{Introduction}

 The metallic spinel  LiV$_2$O$_4$ has attracted much attention
since a heavy fermion behaviour in this material was
reported~\cite{Kondo97,Johnston00, Kondo99}.  Despite  continuous
activity in last years, there is currently no consensus on the
mechanism for formation of heavy fermion quasiparticles in
LiV$_2$O$_4$, and the issue is still under debate~\cite{Fulde04,
Arita07}.

At low temperatures, $T< 30$ K, the spin system of  LiV$_2$O$_4$
exhibits pronounced short-range antiferromagnetic (AFM)
correlations~\cite{Krimmel99,Lee01,Murani04}, but no long-range
magnetic ordering was detected at any measured temperatures. The
geometrical frustration of the pyrochlore lattice of vanadium ions
(in the mixed valence state V$^{3.5+}$) is likely to be a crucial
aspect of the problem. The frustration may suppress  at any $T$ a
long-range ordering of strongly correlated itinerant electrons,
but instead, the system is placed near to a magnetic instability.
The emergence of largely degenerate low lying spin excitations in
the ground state of  LiV$_2$O$_4$ is expected to be responsible
for  low-$T$ properties of this material, including its heavy
fermion behaviour. This appealing picture has been developed in
detail in previous work~\cite{Yushankhai07, Yushankhai08}. It  can
be examined by  considering experimental results obtained by
different techniques, like the nuclear magnetic resonance (NMR)
and the inelastic neutron  scattering (INS), probing  low
frequency spin fluctuations.

The $^7$Li-NMR studies of the spin fluctuation dynamics in
LiV$_2$O$_4$  were reported in a series  of
papers~\cite{Fujiwara98,Mahajan98,Kaps01,Johnston05,Zong08}. For
high temperatures, $T> 60$ K,  the NMR relaxation data have been
successfully explained~\cite{Mahajan98} in terms of a V local
moment formalism, while in the low-T region, $T< 30$ K, an
approach based on the itinerant and strongly correlated electrons
in the paramagnetic metal LiV$_2$O$_4$ is more appropriate.
Low-$T$  measurements  down to $\sim$1K on  samples of high purity
under ambient pressure  reveal  a nearly constant value of the
Knight shift $K$ and  linear $T$ dependence of the nuclear
spin-lattice relaxation rate $T_1^{-1}$ as for normal metals, but
with a very high value of  $(T_1T)^{-1}$.  An
estimate~\cite{Mahajan98} of the Korringa relation $K^2T_1T/S =
R$, where $S= \hbar \gamma_e^2/4\pi k_B\gamma_n^2$, with
$\gamma_e$ and   $\gamma_n$ being the electronic  and nuclear
gyromagnetic ratios respectively, and   $R\approx 0.5$ is less
than unity, indicates~\cite{Moriya63,Narath68} that presumably AFM
fluctuations are dominant in the spin-lattice relaxation at low
$T$.

The NMR measurements on  high purity samples of LiV$_2$O$_4$ under
pressure up to $\approx 5$ GPa  were also reported by K. Fujiwara
{\it et al}~\cite{Fujiwara02,Fujiwara04}. For $T< 10$ K,  the
value of  $(T_1T)^{-1}$  which gives information on the ${\bf q}$
averaged dynamical spin susceptibility  becomes larger on applying
higher pressure and grows with decreasing temperature. At the same
time, the Knight shift $K$ probing only the static unuform
susceptibility $\chi\left({\bf q}=0\right)$ was found to be nearly
temperature independent and insensitive to the pressure above 2
GPa. These results were suggested~\cite{Fujiwara04} to be
indicative of an increase under applying pressure  of AFM spin
correlations at some momenta  ${\bf q\not=0}$ and their enhanced
dominance over those at ${\bf q=0}$.

If  temperature is sufficiently low,  the NMR properties of
LiV$_2$O$_4$ are strongly affected by a small amount of magnetic
defects, $n_{defect}< 1\ $  mol
$\%$~\cite{Kaps01,Johnston05,Zong08}.  The relaxation of the
longitudinal nuclear magnetization versus time is no more  a
single exponential one but described  by a stretched  exponential
function with the characteristic   relaxation rate $1/T_1^{\ast}$
showing a  peak at some temperature $T_{peak}\sim 1$K.  Such a
behaviour was proved~\cite{Zong08} to originate from the spin
freezing of  magnetic defects below $T_{peak}$. With decreasing
$n_{defect}$,  the peak position of   $1/T_1^{\ast}$ is apparently
shifted to lower $T$. Remarkably,  a proper model analysis  of the
low-$T$  NMR data obtained in powder samples of  LiV$_2$O$_4$ with
varying  $n_{defect}< 1$ mol $\%$ has shown~\cite{Zong08} that
relaxation effects due to inhomogeneously distributed magnetic
defects and homogeneous spin fluctuations inherent to magnetically
pure LiV$_2$O$_4$ are separable and thus can be examined
independently. It is worth noting that  in a  single crystal of
LiV$_2$O$_4$ containing a small amount of magnetic impurities or
crystal defects, a somewhat  different behaviour was
observed~\cite{Zong08}, possibly because of a lack of the
separability of the relaxation effects mentioned.

A scenario explaining  the considerable influence of a weak
disorder on the low-$\omega$ spin dynamics detected by the low-$T$
NMR measurements on  LiV$_2$O$_4$ was proposed by Johnston {\it et
al}~\cite{Johnston05},  following a more general consideration
developed in ~\cite{Millis03}.  As already noted, a critical
aspect of the problem is the emergence in the ground state of pure
LiV$_2$O$_4$  of a large number of low lying spin excitations,
implying a proximity of the system  to a magnetic instability.
Inhomogeneously distributed magnetic defects may locally lift the
degeneracy of low lying spin excitations and cause their partial
condensation, thus giving rise to a strong change of spin dynamics
at sufficiently low temperatures. A large degeneracy of strongly
enhanced and slow spin fluctuations in pure LiV$_2$O$_4$  was
confirmed by a combined analysis  of the low temperture INS
data~\cite{Krimmel99,Lee01,Murani04} and the complementary
calculations~\cite{Yushankhai07}  of the dynamic spin
susceptibility   $\chi\left({\bf q},\omega\right)$. In these
calculations performed first at  $T=0$,  the actual electronic
band structure of LiV$_2$O$_4$ obtained in the local-density
approximation is used and effects of strong electron correlations
are treated in the random phase approximation.  As an extension
for finite $T$,  the self-consistent renormalization  (SCR)
theory~\cite{Moriya85} of spin fluctuations was
proved~\cite{Yushankhai08} to be a  helpful tool in explaining the
temperature renormalization of the low-$\omega$ spin fluctuation
dynamics in  LiV$_2$O$_4$  derived from INS
measurements~\cite{Krimmel99,Lee01,Murani04}.

In the present study, the parametrized  SCR theory  is applied to
elucidate the main features of the temperature and pressure
dependences of the spin-lattice relaxation rate observed in the
low-$T$ NMR measurements on  LiV$_2$O$_4$.  As
known~\cite{Moriya85}, the SCR theory offers a phenomenological
description for spin fluctuations in nearly ferro- or
antiferromagnetic itinerant electron systems by taking into
account effects of mode-mode coupling between spin fluctuations
either at  ${\bf q}=0$ or ${\bf q}\not=0$, respectively;  the
latter case is applicable to the paramagnetic spinel LiV$_2$O$_4$.
For details we refer to our recent work~\cite{Yushankhai08}, where
the basic equation of the SCR theory is solved numerically and the
results are compared with INS data for  LiV$_2$O$_4$.  There, the
values of  empirical parameters entering the SCR theory are
estimated to provide the best overall coincidence between the
theory and  INS experiment.

The outline of the paper is as follows. In section 2,  an
expression for the spin-lattice relaxation rate  $T_1^{-1}$ is
derived in terms of the SCR theory and, first, an evolution of
$(T_1T)^{-1}$ down to $\sim 1$ K with increasing pressure is
examined. Next, since the temperature behaviour of  $(T_1T)^{-1}$
down to much lower temperature $\sim 60$mK is available only at
the highest applied pressure of 4.74 GPa, these particular data
deserve a special attention.  We argue, contrary to what K.
Fujiwara {\it et al}  suggested~\cite{Fujiwara04}, that  the
upturn of  $(T_1T)^{-1}$ detected below 0.6 K  is not entirely due
to homogeneous critical AFM spin fluctuations, but more likely is
a signature of an additional nuclear spin relaxation mechanism,
probably  due to the spin freezing of  magnetic defects. Summary
and concluding remarks can be found in section 3.


\section{SCR theory for relaxation rate $1/T_1$  in  LiV$_2$O$_4$ }
\subsection{Background}
The nuclear spin-lattice relaxation rate  due to electronic spin fluctuations is generally given by
\begin{equation}
\frac{1}{T_1} = \frac{2 \gamma_n^2 k_B T}{Ng^2\mu_B^2} \sum_{{\bf q}}|A_{{\bf q}}|^2\frac{\mbox{Im}
\chi\left({\bf q}, \omega_n\right)}{\omega_n},
\label{a1}
\end{equation}
where $A_{\bf q}$ is a ${\bf q}$ dependent effective hyperfine
coupling; in our calculations, the resonance frequency $\omega_n$
will be taken in the limit $\omega_n\to 0$. In  (\ref{a1}),  the
${\bf q}$ summation is over the Brillouin zone (BZ) of the fcc
lattice inherent to the pyrochlore lattice of V atoms in the
spinel structure  LiV$_2$O$_4$; $\chi\left({\bf q}, \omega
\right)$ is the dynamic spin susceptibility calculated per
primitive cell (four V atoms) in units of $(g\mu_B)^2$.
Calculations  suggest~\cite{Yushankhai07}  a  rather peculiar
${\bf q}$-dependence of $\chi\left({\bf q}, \omega \right)$. The
resulting model for a distribution in ${\bf q}$ space of dominant
spin fluctuations was checked~\cite{Yushankhai08} to provide a
firm ground to describe the low-$T$ INS measurements for
LiV$_2$O$_4$. Main features of the model are discussed and used
below to calculate $1/T_1$.

In the low-$T$ limit,  the paramagnetic state of  LiV$_2$O$_4$ is
characterized by strongly enhanced and slow spin fluctuations
occupying a large region in ${\bf q}$ space around the  surface of
a mean radius $|{\bf q}|\simeq Q_c\simeq$ 0.6 $\AA^{-1}$, called
the critical surface~\cite{Yushankhai07} (see Figure 1).  These
strongly degenerate  low-$\omega$ AFM spin fluctuations dominate
over those at smaller ${\bf q}$. Such a rather peculiar
distribution in ${\bf q}$ space of the dominant AFM spin
fluctuations is a consequence of  geometrical frustration of the
pyrochlore lattice of V atoms,  which can be traced
back~\cite{Yushankhai07} to the underlying electronic band
structure and the many sheet Fermi surface of the metallic spinel
LiV$_2$O$_4$.  With increasing $T$ (up to 60 K), the AFM
fluctuations get suppressed while those at the BZ center remain
nearly $T$ independent. As noted in ~\cite{Yushankhai08}, the
observed~\cite{Krimmel99,Lee01,Murani04} warming shift of the
low-$\omega$ integrated INS intensity (from $|{\bf q}|\simeq Q_c$
at $T\to 0$  toward low ${\bf q}$ values at higher temperatures)
does not require any significant temperature renormalization of
$\chi\left({\bf q}, \omega \right)$  at small ${\bf q}$.

\subsection{$1/T_1T$ obtained from  SCR theory}

For the pure  LiV$_2$O$_4$,  two main contributions to the
spin-lattice relaxation rate can be written as
\begin{equation}
 \frac{1}{T_1T}=\left(\frac{1}{T_1T}\right)_{q\sim 0} + \left( \frac{1}{T_1T}\right)_{q\sim Q_c},
\label{a2}
\end{equation}
As discussed later, at $T\to 0$, the second contribution from AFM
spin fluctuations at  $|{\bf q}|\sim Q_c$ is  larger than the
first one coming from  the small ${\bf q}$ spin fluctuations. For
finite but low $T$,  because of a comparatively small variation
with both temperature and  pressure of the Knight shift, and
hence, of the intrinsic uniform susceptibility $\chi\left({\bf
q}=0, \omega=0 \right)$, one expects for the low-$\omega$ and
small ${\bf q}$ spin fluctuations  much weaker temperature and
pressure dependences than those
observed~\cite{Krimmel99,Lee01,Murani04}  for AFM spin
fluctuations. Therefore, in the subsequent analysis the
contribution $\left(1/{T_1T}\right)_{q\sim 0}$  is assumed to be a
constant and treated below as an adjustable parameter of the fit
procedure. In this approximation, we relate the experimentally
observed temperature and pressure dependence of $1/T_1T$ merely to
that of dominant AFM fluctuations around the $Q_c$ surface.

Let us choose a wave vector ${\bf Q_c}$ with the end point lying
on the $Q_c$ surface, and consider its vicinity, ${\bf q}^\prime =
{\bf Q_c}+{\bf q}$, ($|{\bf q}|\ll | {\bf Q_c}|$), as depicted in
Figure 1.  The inverse dynamic spin susceptibility can now be
expanded as
\begin{equation}
\frac{1}{\chi\left({\bf Q}_c+{\bf q}, \omega; T \right)}=\frac{1}{\chi\left({\bf Q}_c; T\right)}
+ A\left(q^{||}\right)^2 + B\left({\bf q}^{\bot}\right)^2 -iC\omega, \label{a3}
\end{equation}
where  $q^{||}$ and ${\bf q}^{\bot}$ are components of ${\bf q}$
parallel and perpendicular to the normal ${\bf Q}_c/ |{\bf Q}_c|$
to the $Q_c$ surface. Since the dynamic spin susceptibility is
considered around ${\bf q}=0$, only the leading term of $C$
independent of ${\bf q}$ is taken in the expansion (\ref{a3}).
This allows us to avoid an increase of the number of parameters in
describing the dynamic spin susceptibility.

\begin{figure}
\includegraphics[width=100mm,height=55mm]{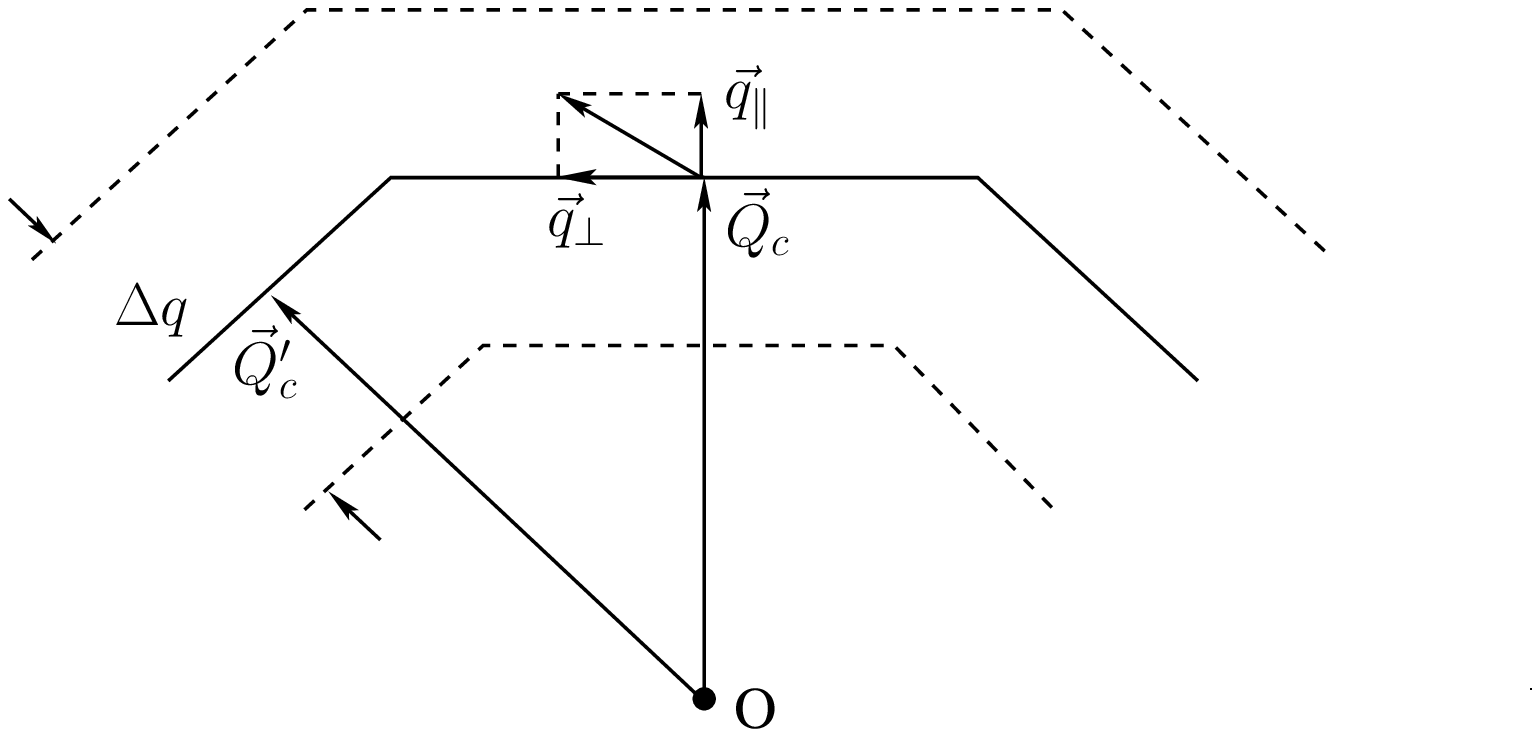}
\caption{ A cross-section in ${\bf q}$ space of the critical $Q_c$
surface ($|{\bf Q_c}|\simeq 0.6 \AA^{-1}$) is shown schematically
by the solid line.  Strongly enhanced antiferromagnetic spin
fluctuations dominating low-$T$ properties of LiV$_2$O$_4$ are
located in a close vicinity (of width $ \Delta q \lesssim 0.2
\AA^{-1}$) to the $Q_c$ surface. The meaning of decomposition
${\bf q}^\prime = {\bf Q_c}+{\bf q}_{||}+{\bf q}_{\bot}$ for an
arbitrary wave vector  ${\bf q}^\prime$ near the $Q_c$ surface is
explained in the text.} \label{fig:Fig.1}
\end{figure}

In  (\ref{a3}),  $A$, $B$ and $C$ are empirical parameters that,
together with $\chi\left({\bf Q}_c; T=0\right)$, can be
estimated~\cite{Yushankhai08} from INS and magnetic measurement
data. The parameters $A,B$ and $C$ are usually taken  to be
$T$-independent in the low-$T$ region where the SCR theory  works
well. Strong anisotropy in ${\bf q}$ space of AFM spin
fluctuations, i.e. $b=B/A\ll 1$, is assumed and
verified~\cite{Yushankhai08}. It is also reasonable using a
spherical approximation to the $Q_c$ surface, implying that
$\chi\left({\bf Q}_c; T\right)$ does not depend on the direction
of ${\bf Q_c}$, i.e.,  $\chi\left({\bf Q}_c; T\right) =
\chi\left(Q_c; T\right)$.

To be close to the standard notation of the SCR
theory~\cite{Moriya85}, we introduce, instead of $A$ and $C$,  the
following parameters (here, $\hbar = k_{B} = g\mu_B=1$ ):
\begin{equation}
T_A=\frac{Aq_B^2}{2},\hspace{5mm}T_0=\frac{Aq_B^2}{2\pi C}, \label{a4}
\end{equation}
where $q_B$ is the effective radius of the BZ boundary given in
terms of the lattice primitive cell volume $v_0$ as
$q_B=\left(6\pi^2/v_0\right)^{1/3}$. Next,  the reduced inverse
susceptibility at $|{\bf q}|= Q_c$ is defined as
\begin{equation}
y_Q\left(T\right)=\frac{1}{2T_A\chi\left(Q_c; T\right)}. \label{a5}
\end{equation}

In this notation, one obtains for $\mbox{Im} \chi\left({\bf q},
\omega_n\right)/\omega_n$ in the limit $\omega_n \to 0$  and up to
a  constant factor the following expression
\begin{equation}
\frac{\mbox{Im}\chi\left({\bf Q}_c+{\bf q}, \omega_n;
T\right)}{\omega_n} \sim \frac{1}{T_0T_A}\frac{1}{\left[
y_Q\left(T\right) + \left(q^{||}/q_B\right)^2  + b\left({\bf
q}^{\bot}/q_B\right)^2\right]^2}. \label{a6}
\end{equation}

When performing  the integration over ${\bf q}$ in (\ref{a1}), two
dimensionless cutoff parameters,  $x_c=(q_c^{\bot}/q_B)^2$  and
$z_c=(q_c^{||}/q_B)$, are introduced. For the former, the
requirement $x_c<1$ is sufficient.  As  the latter cutoff we take
$z_c\simeq 1/2$,  which distinguishes the region of the dominant
AFM spin fluctuations from that near the BZ center. Near the $Q_c$
surface, the hyperfine coupling  is assumed to be a constant
$A_{Q_c}$.  Then the resulting expression for
$\left(1/{T_1T}\right)_{q\sim Q_c}$  reads as
\begin{eqnarray}
\left( \frac{1}{T_1T}\right)_{q\sim Q_c}&=& \frac{3\gamma_n^2\hbar |A_{Q_c}|^2}{\pi k_B T_0 T_A}\left(\frac{Q_c}{q_B}\right)^2\frac{1}{bx_c}\left\{\frac{1}{\sqrt{y_{Q}(T)}}\tan^{-1}\frac{z_c}{\sqrt{y_Q(T)}}\right.\nonumber\\
 &&\left.-\frac{1}{\sqrt{y_Q(T)+bx_c}} \tan^{-1}\frac{z_c}{\sqrt{y_Q(T)+bx_c}}\right\},  \label{a7}
\end{eqnarray}
including again the proper dimensional constants.

\subsection{Basic equation of SCR theory and empirical parameters}
 The reduced inverse susceptibility $y_Q\left(t\right)$, where $t=T/T_0$, obeys the following integral
 equation~\cite{Yushankhai08}:
\begin{equation}
y_Q\left(t\right)=y_Q\left(0\right)+g_Q\int_{0}^{z_c}dz\frac{\phi\left(\left[y_Q\left(t\right) +
z^2 \right]/t\right)-\phi\left(\left[y_Q\left(t\right) + z^2 +bx_c \right]/t\right)}{bx_c /t},
 \label{a8}
\end{equation}
with
\begin{eqnarray}
\phi\left(u\right) &=&\ln  \Gamma\left(u\right)-\left(u-\frac{1}{2}\right)\ln u +u -\frac{1}{2}\ln 2\pi, \label{a9}
\end{eqnarray}
where $\Gamma\left(u\right)$ is the gamma function.

The present  SCR theory includes a set of five parameters which
are now denoted as  $y_Q\left(0\right)$,  $T_A$,  $T_0$,  $g_Q$
and  $bx_c$.  The parameters  $T_A$ and  $T_0$ characterize, at
$T\to 0$, the momentum and frequency  spread of the dominant AFM
spin fluctuations,  $g_Q$  is the effective mode-mode coupling
constant and   $bx_c (\ll 1)$ describes a large anisotropy of the
spin fluctuation dispersion  (\ref{a3})  in ${\bf q}$-space.
  From a fit to INS data, we obtained~\cite{Yushankhai08} the following estimates
\begin{equation}
 T_0\simeq 60 K,  \hspace{2mm} T_A\simeq 220 K,  \hspace{2mm} g_Q=0.16,  \hspace{2mm} bx_c=0.01,
\label{a10}
\end{equation}
and $y_Q\left(0\right)\simeq 0.044$.

The parameter $y_Q\left(0\right)=[2T_A\chi\left(Q_c;
T=0\right)]^{-1}$ is a measure of distance from the magnetic
instability. Following  earlier studies~\cite{Ishikagi96,
Kondo02},  we consider  $y_Q\left(0\right)$ to be the only
pressure dependent parameter, assuming that $y_Q\left(0\right)\to
0$ as the system approaches a quantum critical point with
increasing  pressure. In case of ambient pressure, both the
empirical parameter $y_Q\left(0\right)\simeq 0.044$ and the
solution of equation (\ref{a8}) are taken the same as
in~\cite{Yushankhai08}, thus providing the agreement with INS
data.  Here, two remaining fit parameters for  $1/T_1T$ under
ambient pressure are found to be $\left(1/T_1T\right)_{q\sim
0}\simeq 0.55$ (sec$^{-1}$K$^{-1}$) and $|A_{Q_c}|\simeq 5$ kG.
For  NMR data obtained under the applied pressure,  the only
adjustable parameter $y_Q\left(0\right)$  is estimated by solving
first the equation (\ref{a8}) and inserting the solution
$y_Q\left(T\right)$ into (\ref{a7}).

\subsection{Comparison to experimental data}
The experimental data~\cite{Fujiwara04} for $\left(1/T_1T\right)$
measured for different applied pressure  together with theoretical
curves fitting these data are depicted in Figure 2. A good
agreement between the theoretical results and the available
experimental data is found at least down to 1 K. First we note
that  away from the instability, i.e., $y_Q\left(0\right)>0$, the
SCR theory predicts the Korringa behaviour ($1/T_1T=$ constant) in
the low-$T$ limit. On applying higher pressure (or, as
$y_Q\left(0\right)\to 0$),  $1/T_1T$ becomes larger, while a
temperature range near $T=$0, where the Korringa relation holds,
shrinks and tends to zero at  $y_Q\left(0\right)\to 0$.  These
results of the SCR theory  are in accordance with the  temperature
and pressure dependence of  $\left(1/T_1T\right)$ observed in
LiV$_2$O$_4$.

\begin{figure}
\includegraphics[width=110mm,height=100mm]{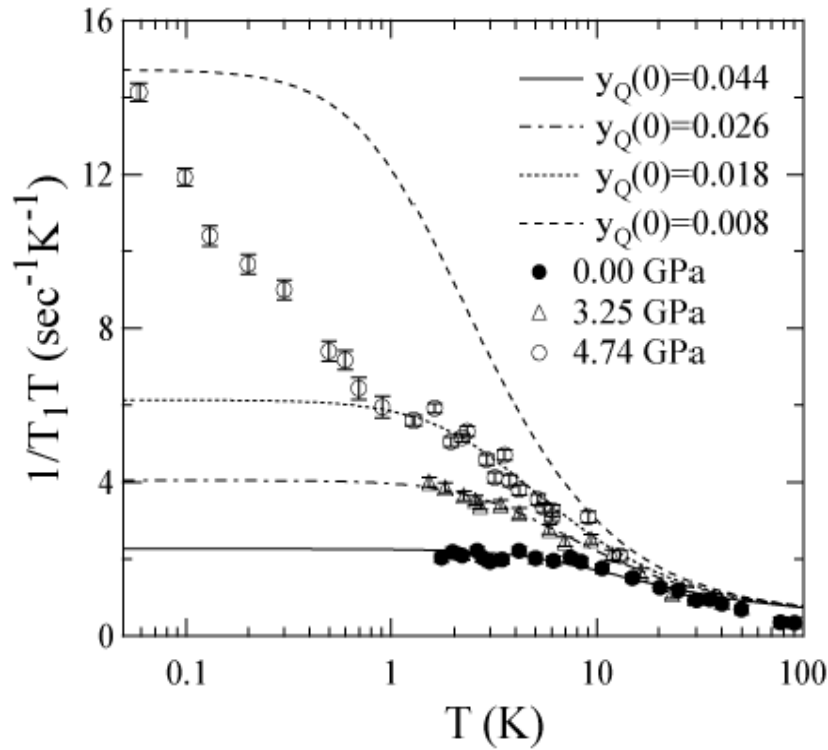}
\caption{ Temperature dependence of  $1/T_1T$  obtained  from
$^7$Li-NMR measurements on powder samples of LiV$_2$O$_4$ under
different applied pressure; the data are taken from K.  Fujiwara
{\it et al} \cite{Fujiwara04}.
Different curves together with corresponding values of the fit parameter
$y_Q\left(0\right)$ represent the calculations based on the SCR theory.\\
}
\label{fig:Fig.2}
\end{figure}

The SCR theory  fails, however,  in giving a quantitative
description of the low-$T$ upturn of $1/T_1T$ detected below 0.6 K
under the highest applied pressure of 4.74 GPa. For instance, as
seen from Figure 2, the parameter $y_Q\left(0\right)=$0.018 fits
well in describing the experiment down to $\sim 1$K,  while for
0$<y_Q\left(0\right)<$0.018 a strong deviation in the whole
temperature range is found (see, for instance, the upper
theoretical curve corresponding to $y_Q\left(0\right)=$0.008). It
is worth emphasizing that just at the quantum critical point,
$y_Q\left(0\right)$=0,  the equation (\ref{a8}) leads to the power
law  $\left(1/T_1T\right)\sim T^{-3/4}$,  in agreement with the
earlier SCR theory predictions~\cite{Ishikagi96,Kondo02} for the
quantum critical behaviour of   $\left(1/T_1T\right)$ around the
AFM instability in three-dimensional metals. In contrast, the
power law $\left(1/T_1T\right)\sim T^{-1/3}$ derived from the
experimentally observed  behaviour of  $\left(1/T_1T\right)$ in a
wide temperature range, 0.1 K $< T < $ 10 K, was reported in
~\cite{Fujiwara04}.

The discrepancy can be clearly explained by taking into account
that the SCR theory is invoked to describe homogeneous spin
fluctuations in a pure LiV$_2$O$_4$, while a certain  amount of
crystal defects and/or  magnetic impurities  unavoidably present
in the measured powder samples of  LiV$_2$O$_4$ may contribute to
the nuclear spin relaxation as well. Actually, when approaching  a
magnetic instability and softening of largely degenerate low lying
spin fluctuations, the system becomes very susceptible to  weak
perturbations including, for instance, magnetic defects.  As
known~\cite{Kaps01,Johnston05,Zong08},   the cooperative
properties of the paramagnetic  LiV$_2$O$_4$ detected by NMR under
ambient pressure can be drastically changed at sufficiently low
temperatures $T\sim 1$ K due to a small amount of magnetic
defects. We suggest that the low-$T$ upturn of
$\left(1/T_1T\right)$ starting at $T_{upturn}\approx$ 0.6 K is
more likely to be a manifestation of the onset of spin freezing of
magnetic defects. If so,  this calls for a complete reexamination
of the experimental data below $T_{upturn}$, as done, for
instance, in ~\cite{Zong08} by using a stretched exponential
description for the nuclear spin relaxation. We expect that
measurement of $\left(1/T_1T\right)$ on a sample with the same
quality, but under the ambient pressure,  would reveal the onset
of  spin freezing  at a lower temperature, $T_{upturn} < 0.6$ K,
because of a larger distance from the instability.

Finally, it is helpful to discuss the $T$-dependence of $1/T_1$
predicted by the SCR theory for higher temperatures, $T > T_0$.
For instance, a typical behaviour $1/T_1\sim \sqrt{T}$ is expected
in the paramagnetic state of nearly and weakly AFM
three-dimensional metals ~\cite{Moriya85,Ishikagi96}. By extending
formally the present SCR theory to the region $T > T_0$, we obtain
$1/T_1 \sim T^{\alpha}$ with $\alpha >1/2$, because of the SCR
equation (\ref{a8}) reflecting an effective quasi
one-dimensionality due to the strong anisotropy in ${\bf q}$ space
of AFM spin fluctuations in our model. We recall, however, that
the present SCR theory works well up to 40 K, which is below $T_0$
characteristic to LiV$_2$O$_4$, as evidenced from the comparison
of theoretical results with  INS data ~\cite{Yushankhai08}.
Actually, for $T > T_0$, the AFM fluctuations at $|{\bf q}|\simeq
Q_c$ are suppressed and no more distinguished from those at other
wave vectors in BZ; the system enters a spin localized regime
compatible with the Curie-Weiss behavior of $\chi({\bf q}=0)$
observed in LiV$_2$O$_4$ for $T > 60$ K.
\section{Summary and conclusions}
AFM spin fluctuations located in a large region of ${\bf q}$ space
around the critical surface of a mean radius  $|{\bf q}|\simeq
Q_c\simeq$ 0.6 $\AA^{-1}$ dominate the low-$T$ properties of the
paramagnetic spinel  LiV$_2$O$_4$. A parametrized self-consistent
renormalization theory of the AFM spin fluctuations was developed
and applied to describe  temperature and pressure dependences of
the low-$T$ spin-lattice relaxation rate $1/T_1$ in this material.
Most of the empirical parameters entering the present  SCR theory
have been estimated  earlier  and kept fixed in the present study.
To simulate pressure effects in our calculations, the inverse
static spin susceptibility $\chi^{-1}\left(Q_c\right)$ is
considered to be the only fit parameter depending on the applied
pressure.

Comparison between  NMR data and the calculated results has shown
that the SCR theory is able to describe correctly the development
of AFM spin fluctuations  as the paramagnetic metallic state of
LiV$_2$O$_4$ approaches a magnetic instability under the applied
pressure. In particular, we concluded that up to the highest
applied pressure of 4.74 GPa, the spin system is still away from
the instability. In this case,  the SCR theory predicts the
Korringa behaviour ($1/T_1T=$ constant) in a narrow temperature
range near $T=$ 0; on applying higher pressure,  a constant value
$1/T_1T$ becomes larger and  a temperature range near $T=$ 0,
where the Korringa relation holds, shrinks. The theoretical
results were shown to be in a good agreement with experimental
data for $1/T_1T$  in  wide ranges of temperature and pressure. A
deviation from the theoretical prediction   was interpreted as a
signature of changing below  the characteristic temperature
$T_{upturn}$ ($<$ 1 K) of the spin fluctuation dynamics. If for
$T>T_{upturn}$ the nuclear spin relaxation is determined by
homogeneous AFM spin fluctuations inherent to a magnetically pure
LiV$_2$O$_4$,  then below  $T_{upturn}$  cooperative properties of
the detected paramagnetic state are strongly affected by coupling
of spin fluctuations  to magnetic defects   in the measured sample
of LiV$_2$O$_4$. A detailed mechanism describing this coupling and
the magnetic defect interactions in   LiV$_2$O$_4$ remains a
challenge~\cite{Johnston05,Zong08} for a further work.

\ack
Discussions  with P  Fulde,  J  Schmalian and T  Vojta  are gratefully acknowledged.
This work was supported  by Deutsche Forschungsgemeinschaft,  the project  SFB 463.

\end{document}